\documentclass[graybox, envcountchap]{svmult}

\usepackage{mathptmx}        
\usepackage{amsmath}
\usepackage{amssymb}
\usepackage{color}
\usepackage{helvet}          
\usepackage{courier}         
\usepackage{dirtree}

\usepackage{makeidx}        
\usepackage{graphicx}        
\usepackage{subfig}

\usepackage{multicol}        
\usepackage[bottom]{footmisc}

\usepackage{hyperref}        
\hypersetup{colorlinks=true,urlcolor=blue}

\usepackage[misc]{ifsym}
\usepackage{xspace}
\usepackage{tabularx}
\usepackage{array}

\makeindex             

\newcommand\kerrc{{\tt kerrC}\xspace}
\newcommand\sherpa{{\tt Sherpa}\xspace}
\newcommand\xtrack{{\tt xTrack}\xspace}
\newcommand\Teff{T_{\rm eff}}
\newcommand\Tcol{T_{\rm col}}
\newcommand\Ei{E_{\rm i}}
\newcommand\fh{f_{\rm h}}
\newcommand\tauC{\tau_{\rm C}}
\newcommand\TC{T_{\rm C}}
\newcommand\rg{r_{\rm g}}
\newcommand\xillver{{\tt xillver}\xspace}

\begin{document}


\title{Modeling of the X-ray Flux and Polarization Energy Spectra of Black Hole X-Ray Binaries with the KerrC Code}
\author{Kun Hu and Henric Krawczynski}
\institute{Kun Hu (\Letter) \at Washington University in St. Louis, St. Louis, MO, USA, \email{hkun@wustl.edu}
\and Henric Krawczynski \at Washington University in St. Louis, St. Louis, MO, USA, \email{krawcz@wustl.edu}}
%
%
\maketitle

\abstract{In this chapter, we introduce the \kerrc model, an advanced framework designed for the analysis of combined spectral and polarimetric X-ray emission from black hole systems. The model is built upon an extensive dataset of ray-tracing Monte Carlo simulations that compute the radiative transfer of X-rays in the strongly curved spacetime near the event horizon of a Kerr black hole. These simulations incorporate thermal seed emission from a geometrically thin, optically thick accretion disk, along with the subsequent Comptonization processes within a hot corona of variable geometry and physical properties. Crucially, the reflection of both primary disk and Comptonized photons is treated self-consistently utilizing the \xillver reflection tables, and the polarization 4-vectors of the photons are rigorously transported throughout the simulation to preserve full polarization information. The code tracks photons forward in time, allowing for multiple photon scatterings in the corona and off the accretion disk. We detail the underlying mechanics of the ray-tracing simulations and the fitting models, and discuss the applications of \kerrc in leveraging X-ray polarization observations to constrain black hole spin and coronal geometry.}



\section{Introduction}

Black hole X-ray binaries (BHXRBs) are dynamic astrophysical systems consisting of a stellar mass black hole accreting matter from a companion star. The primary sources of X-ray emission in these systems are a geometrically thin, optically thick accretion disk that emits thermal radiation \cite{Shakura:1972te,Novikov:1973kta}, and a tenuous, optically thin electron plasma known as the corona, which produces a non-thermal hard X-ray power-law tail via inverse Compton scattering of the disk photons \cite{1980A&A....86..121S,1991ApJ...380L..51H}. BHXRBs are observed to transition between distinct spectral states, primarily determined by their hardness ratios \cite{2006ARA&A..44...49R,2004MNRAS.355.1105F}. The high soft state is characterized by high luminosity and a spectrum dominated by the thermal disk emission. Whereas in the low hard state, the luminosity is lower and the spectrum is dominated by a non-thermal Comptonized coronal component. The low hard state is often accompanied by the launching of relativistic radio jets. Revealing and modeling the physical interplay between the disk, the corona, and the jet remains a central goal of high-energy astrophysics.

The observational landscape of BHXRBs is rapidly advancing through a combination of high-resolution spectroscopy and the dawn of high-sensitivity X-ray polarimetry. For decades, foundational spectral constraints on accretion disk, relativistic reflection and the coronal properties have been driven by flagship observatories such as \emph{XMM-Newton} \cite{2001A&A...365L...1J}, \emph{NICER} \cite{2016SPIE.9905E..1HG}, and \emph{NuSTAR} \cite{2013ApJ...770..103H}. Building on this rich spectral legacy, missions such as \emph{XRISM} provide unprecedented spectral resolution, allowing for precise measurements of Fe reflection features and outflow kinematics \cite{2020arXiv200304962X,2025PASJ...77.1210Y,2025ApJ...995L..12D}. Simultaneously, X-ray polarimetry has opened a new observational window. The \emph{IXPE} mission has successfully measured the 2--8 keV linear polarization of several BHXRBs, offering direct geometric constraints on the corona and the inner accretion flow \cite{2022JATIS...8b6002W,2022Sci...378..650K,2024ApJ...969L..30S}. At higher energies, balloon-borne missions like \emph{XL-Calibur} target hard X-ray polarization (19--64 keV) to scrutinize the coronal emission without the need to disentangle disk and coronal emission \cite{2021APh...12602529A,2025ApJ...994...37A}. Upcoming satellite missions such as \emph{COSI} will measure the MeV gamma-ray spectrum and polarization, providing critical insights into jet launching and emission mechanisms \cite{2024icrc.confE.745T}.


To interpret these complex multi-wavelength and polarimetric observations, theoretical frameworks must accurately compute radiative transfer in the strongly curved spacetime immediately surrounding the black hole. Over the past two decades, spectral fitting models have been extensively used to measure black hole spin and map the inner disk. These models generally include a multi-temperature disk emission model such as {\tt kerrbb} \cite{2005ApJS..157..335L}, a Comptonized power-law component \cite{1996MNRAS.283..193Z}, and models for relativistic reflection features such as {\tt xillver} or {\tt relxill} \cite{2010ApJ...718..695G,2013ApJ...768..146G,2014ApJ...782...76G}. 
However, these traditional models were designed exclusively for spectral analysis and rely on highly simplified assumptions regarding the corona. They typically approximate the coronal emission as a power law originating from an idealized compact lamppost corona. More importantly, these models do not track the polarization of the emission, therefore they are unable to leverage the polarimetric constraints from missions such as \emph{IXPE} 
and \emph{XL-Calibur} to diagnose the effects of returning radiation and constrain coronal geometries \cite{2009ApJ...701.1175S}.

The \kerrc framework provides a comprehensive, self-consistent tool designed for the joint analysis of spectral and polarimetric data \cite{2022ApJ...934....4K}. To achieve this, \kerrc utilizes extensive Monte Carlo ray-tracing datasets generated by the \xtrack code, which rigorously tracks the full polarization 4-vector of photons as they propagate the strongly curved spacetime near the black hole \cite{2012ApJ...754..133K}. A major strength of \kerrc is its highly accurate treatment of the underlying microphysics, including the fully polarization-dependent Compton scattering cross sections with hot plasma and a precise description of partially ionized disk reflection utilizing the \xillver tables. It also overcomes the limitations of idealized point-source approximation of the corona and enables testing and fitting data against a variety of more realistic corona geometries.
By building upon these robust physical foundations, \kerrc provides the opportunity to fully leverage X-ray polarimetry to constrain both black hole spin and the three-dimensional structure of the corona.

Various other codes have been developed with a similar goal. To our knowledge, \kerrc is the only Monte Carlo code that includes Comptonizing coronae and allows for fitting actual data (as it is based on a large library of simulated models). Furthermore, the code calculates the intensity and energy spectrum of the coronal and returning emission on the fly while fitting actual data and uses them to choose the corresponding \xillver model to reflect the emission. 
{\tt KerrC} adopts elements of the ray-tracing scheme described in \cite{2012ApJ...745....1P}. The earliest version of the code implemented the disk reflection using Chandrasekhar's results in a similar way to \cite{2009ApJ...701.1175S}.
Other notable codes include the
MONK ray-tracing code \cite{2019ApJ...875..148Z},
and the Panduratta radiation transport code \cite{2013ApJ...777...11S,2019ApJ...873...71K,2026arXiv260103349N}. The effects of line emission and absorption has been discussed in
\cite{2021MNRAS.501.3393T}.
\section{Ray-tracing Simulation}
\label{sec:ray_tracing_simulation}

\subsection{Overall Architecture}
\label{subsec:overall_archi}

{\tt XTrack} is a Monte Carlo ray-tracing code to compute radiative transfer in the strong-gravity environment near black holes. It generates a database of different black hole configurations for \kerrc spectro-polarimetric fittings. The \xtrack code adopts a forward-modeling strategy: radiation is followed from its emission site, through curved spacetime and interactions with the surrounding plasma, to a distant observer.
The code is fully relativistic and treats polarization self-consistently. It evolves the polarization four-vector via parallel transport along null geodesics and models Compton scattering in the corona using a Fano-matrix formalism that retains the complete polarization dependence together with the Klein–Nishina suppression. In addition, \xtrack incorporates returning and reflected radiation with a configurable reflection treatment that switches between a fully ionized Chandrasekhar reflection model and the more detailed \xillver reflection tables. 

The \xtrack code treats the radiative transfer of ``photon packets'' which represent a statistical ensemble of real photons that share the same physical properties. In the simulation, the position vector $x^{\mu}$, momentum vector $k^{\mu}$, and polarization vector $f^{\mu}$ of the photon packet are tracked during the propagation, and a weighting factor $w$ is also assigned. The Boyer-Lindquist (BL) coordinates of the Kerr metric are adopted for the photon transport along null geodesics, while scattering and reflection are treated in the local rest frame of plasma with orthonormal bases.

After a photon is injected from the disk in the local Keplerian frame, the momentum and the polarization vectors are transformed back into the global coordinate system. The photon then propagates following a null geodesic, while its polarization vector is evolved via parallel transport along the same trajectory.
We explicitly integrate the polarization transport equation rather than exploiting the conserved Penrose-Walker constant \cite{Walker:1970un,1977Natur.269..128C}. This choice preserves the generality of the implementation and allows the application of the ray-tracing calculation beyond the Kerr metric. 
In a stationary, axisymmetric spacetime, the geodesic equations are conveniently formulated using the conserved quantities associated with the timelike and azimuthal Killing vectors, i.e., the conserved energy $E_{\gamma} =-k_t =-g_{tt} k^t -g_{t\phi} k^{\phi}$ and angular momentum $L_{\gamma} =  k_{\phi}=g_{\phi\phi}k^{\phi}+g_{t\phi} k^t$. Using these constants of motion and the metric coefficients, the equation of motion can be written in first-order form as
\begin{eqnarray}
    \frac{dt}{d\lambda'} &=&  \frac{-g_{\phi\phi}-b g_{t\phi}}{D}  \nonumber\\
    \frac{d\phi}{d\lambda'}&=&  \frac{b g_{tt} +g_{t\phi}}{D} \nonumber\\
     \frac{dr}{d\lambda'}&=&  \frac{k^r}{E_\gamma}  \nonumber\\
    \frac{d\theta}{d\lambda'}&=&   \frac{k^\theta}{E_\gamma}  \nonumber\\
    \text{with}\quad   b=L_{\gamma}/E_{\gamma}, && \text{and} \quad D = g_{\phi\phi} g_{tt}-g_{t\phi}^{\,\,\,2},
\end{eqnarray}
where $ \lambda' = E_\gamma \lambda$ is the rescaled affine parameter. The momentum and polarization vectors are parallel transported through
\begin{eqnarray}
    \frac{d k^{\mu}}{d\lambda'}\,&=&\,
-\Gamma^{\mu}_{\,\,\,\sigma\nu}
k^{\sigma}k^{\nu}/E_\gamma, \nonumber\\
    \frac{d f^{\mu}}{d\lambda'}\,&=&\,
-\Gamma^{\mu}_{\,\,\,\sigma\nu}
f^{\sigma}k^{\nu}/E_\gamma.
\end{eqnarray}
In the \xtrack simulation, the conserved energy $E_\gamma$ and angular momentum $L_\gamma$ are specified when the photon energy and momentum are initialized at the emission site. We then integrate the geodesic equations of motion together with the parallel-transport equations for $k^{\mu}$ and $f^\mu$ using an adaptive step-size Cash-Karp integrator (a fifth-order Runge-Kutta scheme) \cite{Cash:1990jdx}. The step size is adjusted dynamically: it is automatically reduced whenever the optical depth accumulated in a single step exceeds a prescribed threshold (by default 2\% of the total optical depth) and it is further constrained so that the integration steps terminate exactly at region interfaces, preventing a photon from crossing boundaries (e.g., entering or leaving the corona) within a single step. After each step, the normalization and orthogonality conditions $k^2=0$,  $k\cdot f=0$, and $f^2=1$ are checked to ensure the integration accuracy. Photons are followed until they either reach within 1.02 times the radius of the event horizon or reach a distant observer at $10,000\, r_{\rm g}$, with $r_{\rm g}=GM/c^2$ being the gravitational radius. Here, $G$ is the gravitational constant, $M$ is the black hole mass, and $c$ is the speed of light. For the escaped photons, the momentum and polarization vectors are transformed into the frame of a coordinate-stationary observer, and the resulting quantities are recorded for each photon.

\subsection{Disk Emission and Disk Reflection}

The \xtrack simulation assumes that the thermal radiation originates from a geometrically thin but optically thick accretion disk \cite{Shakura:1972te,Novikov:1973kta,Page:1974he}. The disk emissivity is prescribed using a Novikov–Thorne thin-disk model, in which viscous stresses transport angular momentum and the associated torque dissipates orbital energy as heat that is radiated away from the disk. Specifically, the local radiative flux $F(r,a)$ as a function of radius $r$ and black hole spin $a$ is numerically integrated using Eqns.~(11) and (12) of Page \& Thorne (1974) \cite{Page:1974he}, i.e., 
\begin{equation}
    F(r,a)\,=\,\frac{\dot{M}_0}{4\pi}e^{-(\nu+\psi+\mu)} f(r,a),
\end{equation}
with 
\begin{equation}
    f(r,a)\,=\,
\frac{-p^{t}_{\,\,,r}}{p_{\phi}}\,
\int_{r_{\rm ISCO}}^{r}\,
\frac{p_{\phi,r}}{p^{t}}\,dr.
\end{equation}
Here $\nu$, $\psi$, $\mu$ characterize a general axially symmetric metric in the equatorial plane (see Eqn.~1 in Page \& Thorne 1974), $\dot{M}_0$ is the mass accretion rate, and $p^\mu$ is the 4-momentum of the disk plasma. 
Although Page \& Thorne (1974) provide a closed-form analytic expression for 
$F(r,a)$ in the Kerr metric (their Eqn. 15), we evaluate the underlying integral numerically so the same implementation can be applied to alternative spacetime metrics.
The local flux is then converted into the local temperature through
\begin{equation}
    \Teff(r,a)=\left[\frac{F(r,a)}{\sigma_{\rm SB}}\right]^{1/4}
\end{equation}
where $\sigma_{\rm SB}$ is the Stefan-Boltzmann constant. A hardening factor $f_{\rm h}=1.8$ is multiplied by $\Teff$ to obtain the color-corrected temperature, $\Tcol = \fh\Teff$, for the diluted Planck spectrum from the accretion disk \cite{1995ApJ...445..780S}, i.e.,
\begin{equation}
    F_\nu = \frac{1}{{\fh}^4}\pi B_\nu(\nu,\Tcol),
\end{equation}
where $B_\nu$ is the Planck function. 

The photon energy emitted from each disk radial bin is drawn from a Planck spectrum evaluated at the color-corrected temperature $\Tcol$. We exploit the series expansion 
\begin{equation}
    \frac{x^2}{e^x-1} = \sum_{n=1}^{\infty} x^2 e^{-nx}
    =\sum_{n=1}^{\infty}\frac{2}{n^3}\text{Gamma}(x|3,n).
\end{equation}
Here $\text{Gamma}(x|3,n)=(1/2)n^3x^2e^{-nx}$ is the normalized probability density function of the gamma distribution with a shape of 3 and a rate of $n$. Therefore the Planck distribution can be expressed as a weighted sum of gamma distributions with different rates $n$. The photon ensemble can be sampled using a two-stage procedure: we first draw an integer $n \ge 1$ from a Zipf distribution $P(n)\propto n^{-3}$, then draw a continuous variate $G$ from the gamma distribution. The dimensionless photon energy is then obtained from the ratio $G/n$, and the initial energy in the local rest frame is $\Ei = (G/n)k\Tcol$.

 Rather than explicitly simulating black holes across a wide grid of masses and luminosities (or equivalent mass accretion rates), we leverage the thermodynamic rescaling relations of the disk temperature profile established in \cite{Page:1974he}. By fixing the baseline black hole mass in the simulations at $10 M_{\odot}$, we can efficiently cover the parameter space by introducing a temperature scaling factor, 
\begin{eqnarray}
    \sigma(\dot{M},M) = \left(\frac{\dot{M}}{10\%\dot{M}_{\rm Edd}}\right)^{1/4}\left(\frac{M}{10M_{\rm \odot}}\right)^{-1/2}. \label{eqn:sigma}
\end{eqnarray}
Here $\dot{M}_{\rm Edd}$ is the Eddington accretion rate, and $M_{\rm \odot}$ is the solar mass. We run the \kerrc model for a range of $\sigma$ values using scaled disk temperature profiles with 
\begin{eqnarray}
    T(\dot{M},M) =\sigma(\dot{M},M)\times T(10\%\dot{M}_{\rm Edd},10M_{\rm \odot}). \label{eqn:temperature}
\end{eqnarray}
During spectral fitting, when the {\tt kerrC} model is called, 
we interpolate the energy spectra simulated with the nearest $\sigma$ values.

Thermal photons are launched from the disk surface isotropically in the local Keplerian rest frame, restricted to the upper hemisphere $k^\theta\le0$. The initial momentum vector $k^{\hat{\mu}}$ is normalized so that $k^{\hat{t}} = |k^{\hat{i}}| = \Ei$. 
The angular dependence of the emergent intensity is then accounted for by assigning each photon a direction-dependent weighting factor using the limb-darkening information tabulated in Table XXIV of Chandrasekhar (1960) \cite{1960ratr.book.....C}, thereby capturing the anisotropy expected from a scattering-dominated, optically thick atmosphere. 
An extra factor of $\cos{\theta}$ is applied to the intensity table to take into account the projection effect of the oblique beam. 
The initial polarization vector $f^{\hat{\mu}}$ is set in the same local frame, with $f^{\hat{t}}=0$ and the spatial components of $f^{\hat{\mu}}$ parallel to the disk plane and perpendicular to the photon propagation direction. The polarization fraction is also assigned consistently with Chandrasekhar's Table XXIV for the corresponding emission angle.

In the \xtrack simulation, a fraction of the emitted photons will inevitably strike the accretion disk again, either bent back by strong gravitational lensing or redirected via scattering within the corona, resulting in the irradiation of the disk. The code models the interaction using Chandrasekhar's analytic solution for a semi-infinite scattering atmosphere. When a photon hits the disk and a reflection event is triggered, a random outgoing direction is selected for the reflected photon. It then assigns a statistical weighting factor based on Table XXV of \cite{1960ratr.book.....C} that accurately reflects the physical probability of the photon scattering into that chosen direction. The photon's new parameters are calculated to properly capture its new polarization state. 

\begin{figure}
    \centering
    \includegraphics[width=0.32\linewidth]{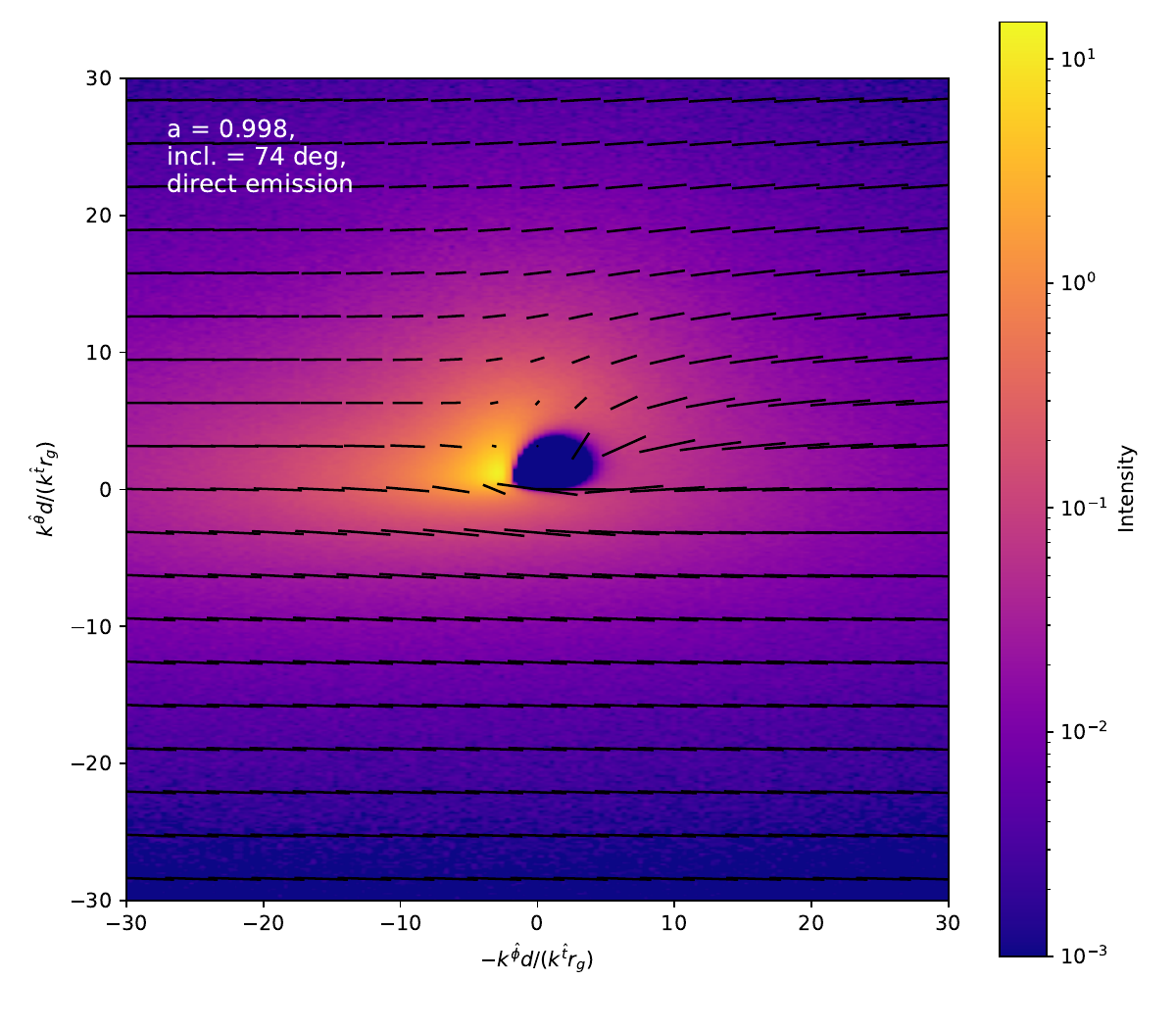}
    \includegraphics[width=0.32\linewidth]{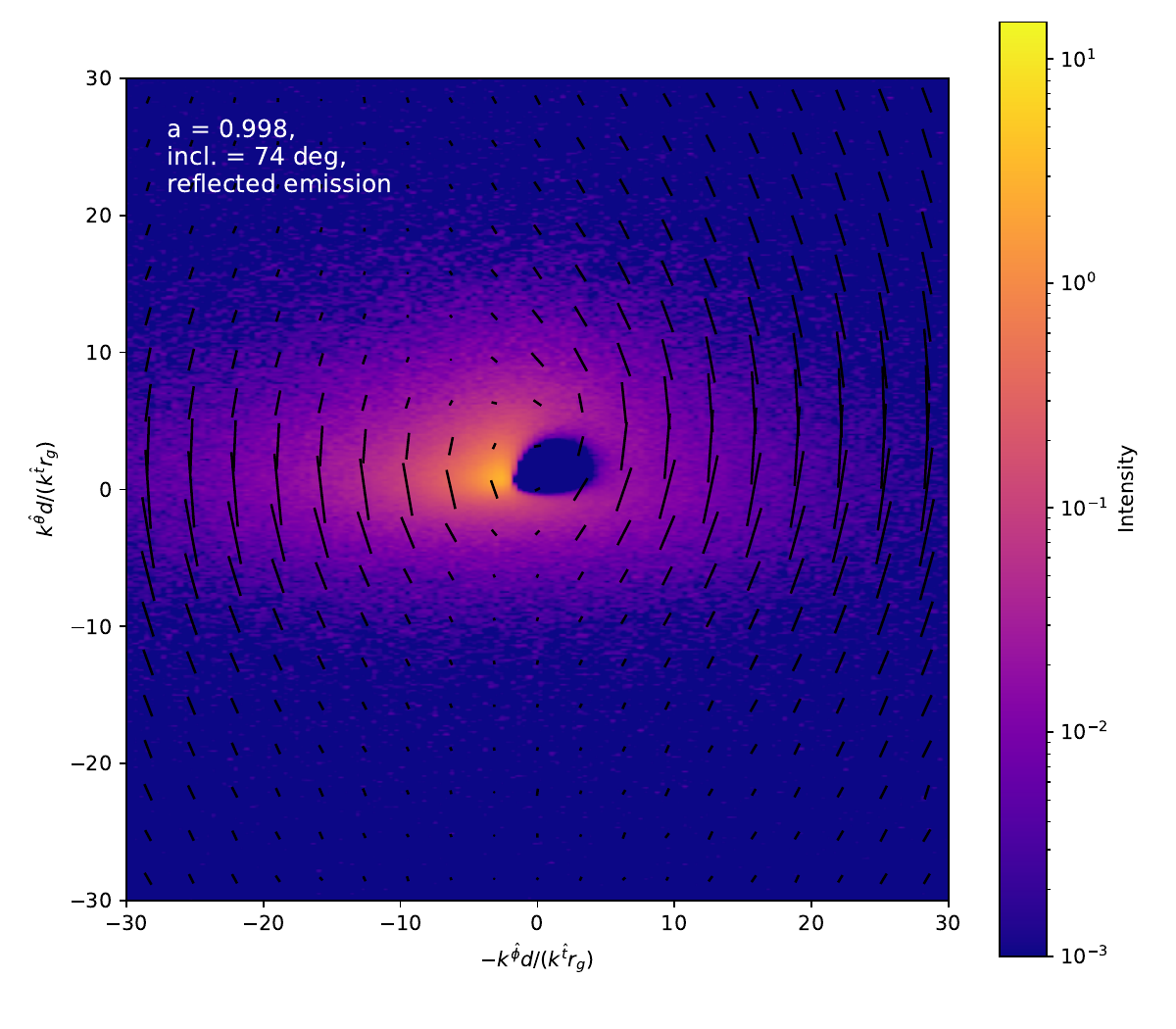}
    \includegraphics[width=0.32\linewidth]{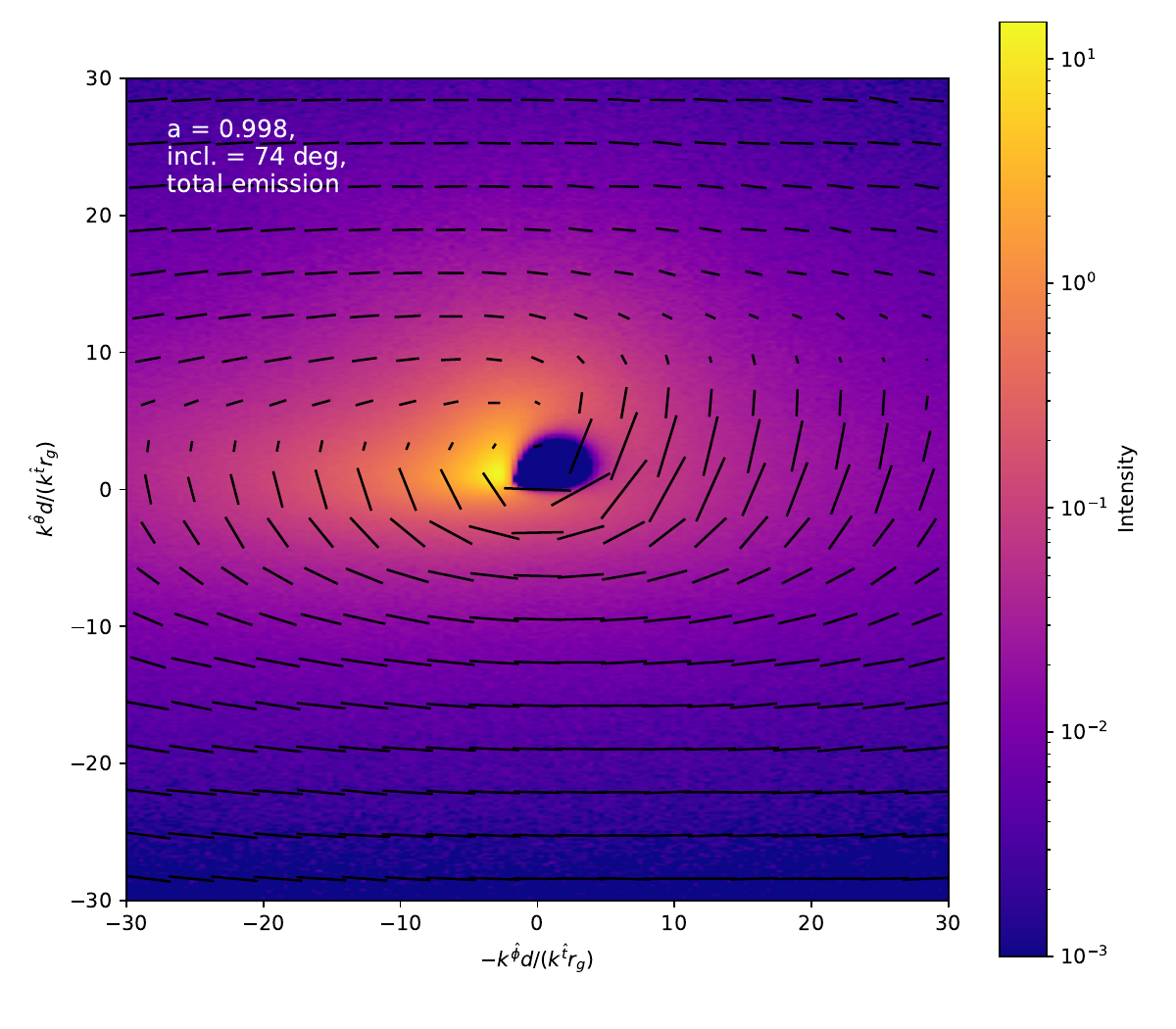}
    \caption{Simulated \xtrack intensity heatmaps for an accreting Kerr black hole with a spin parameter of $a = 0.998$ and an inclination angle of 74 degrees. The temperature scaling factor $\sigma$ is fixed at 0.75. The three panels show the direct, reflected, and the total emission, respectively. The color scale shows the intensity on a logarithmic scale, and the short black bars indicate the polarization direction and relative strength.}
    \label{fig:heatmap}
\end{figure}

The Chandrasekhar model fundamentally depicts reflection from a fully ionized disk, meaning a comprehensive radiative transfer model like \xillver must be used to capture the detailed emission line structures arising from a partially ionized disk. 
To facilitate a self-consistent treatment of reflected radiation, each simulation run outputs three distinct datasets in HDF5 format. These datasets categorize the photon trajectories into direct emission data (photons that reach the observer without interacting with the accretion disk again), reflected emission data (photons that reach the observer after striking the disk at least once), and disk data (a record of all photons that hit the disk surface). 

In Figure \ref{fig:heatmap}, we present heatmaps illustrating the direct, reflected, and total emission from a Kerr black hole ($a=0.998$, $74^\circ$ inclination) generated via the \xtrack simulations. The $x$ and $y$ axes represent the projected distance in units of $\rg$, and the disk temperature scaling factor $\sigma$ is fixed at 0.75. The images feature a distinct asymmetric brightness distribution and a central shadow, characteristic of relativistic beaming in the inner accretion flow and the photon sphere. The local polarization state is mapped using superimposed black lines, where the angle and length of each line correspond to the polarization direction and the relative polarization fraction, respectively. The direct emission is characterized by polarization parallel to the disk plane, consistent with the Chandrasekhar model. Conversely, the reflected photons are mostly polarized perpendicular to the disk because the scattering plane is generally orthogonal to the disk. Notably, while the direct emission contributes the majority of the total flux, the reflected component can strongly influence the polarization of the total emission, allowing perpendicular polarization to dominate the combined signal (see panel 3).

\subsection{Comptonization in the Corona}

\begin{figure}[b]
\centering

\begin{minipage}[c]{0.32\linewidth}
    \centering
    \includegraphics[width=0.95\linewidth]{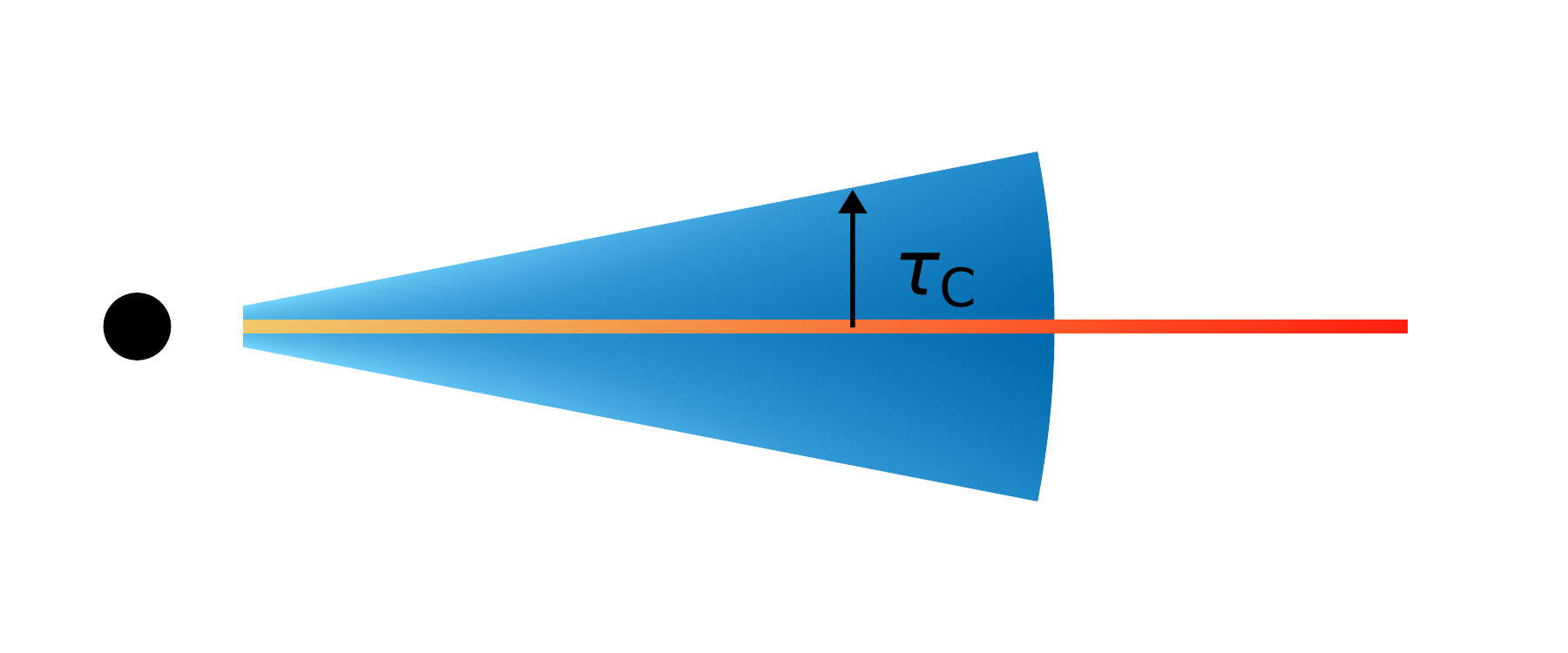}
\end{minipage}\hfill
\begin{minipage}[c]{0.32\linewidth}
    \centering
    \includegraphics[width=1.20\linewidth]{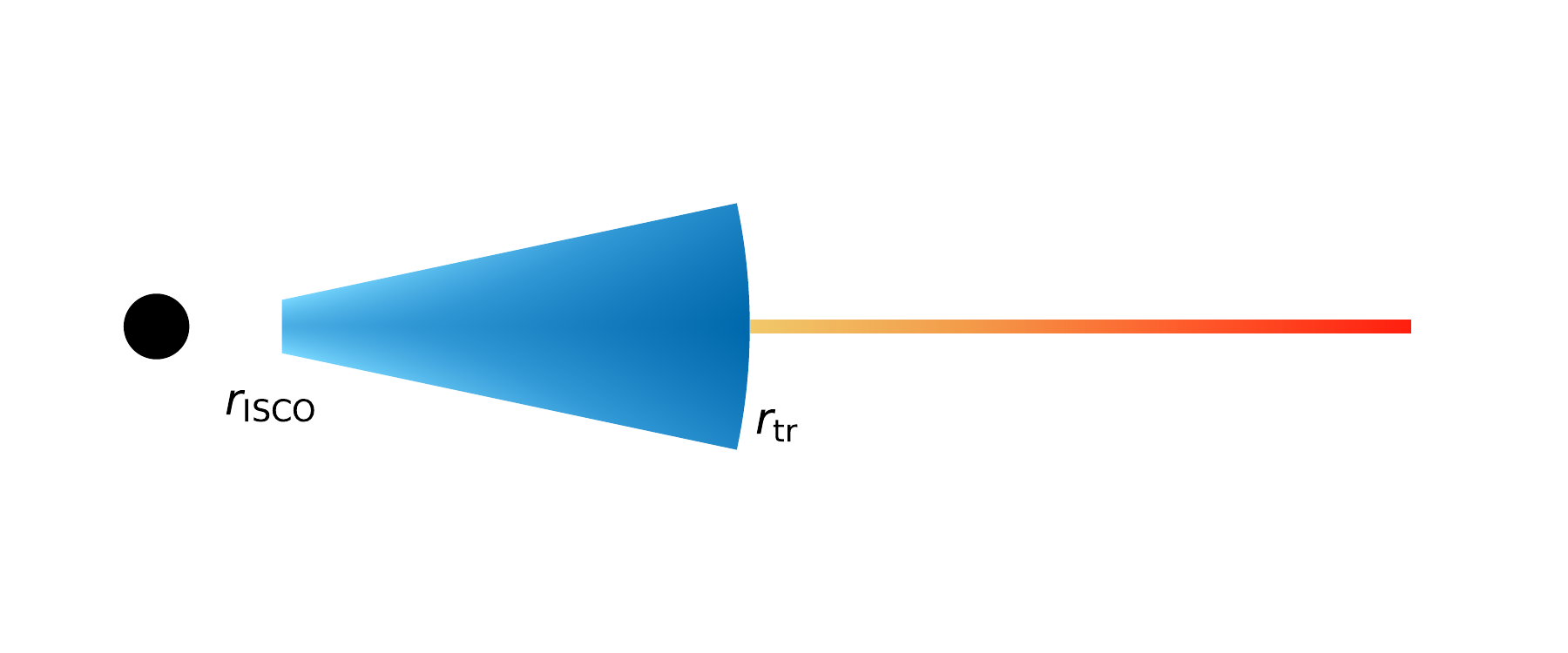}
\end{minipage}\hfill
\begin{minipage}[c]{0.32\linewidth}
    \centering
    \includegraphics[width=0.82\linewidth]{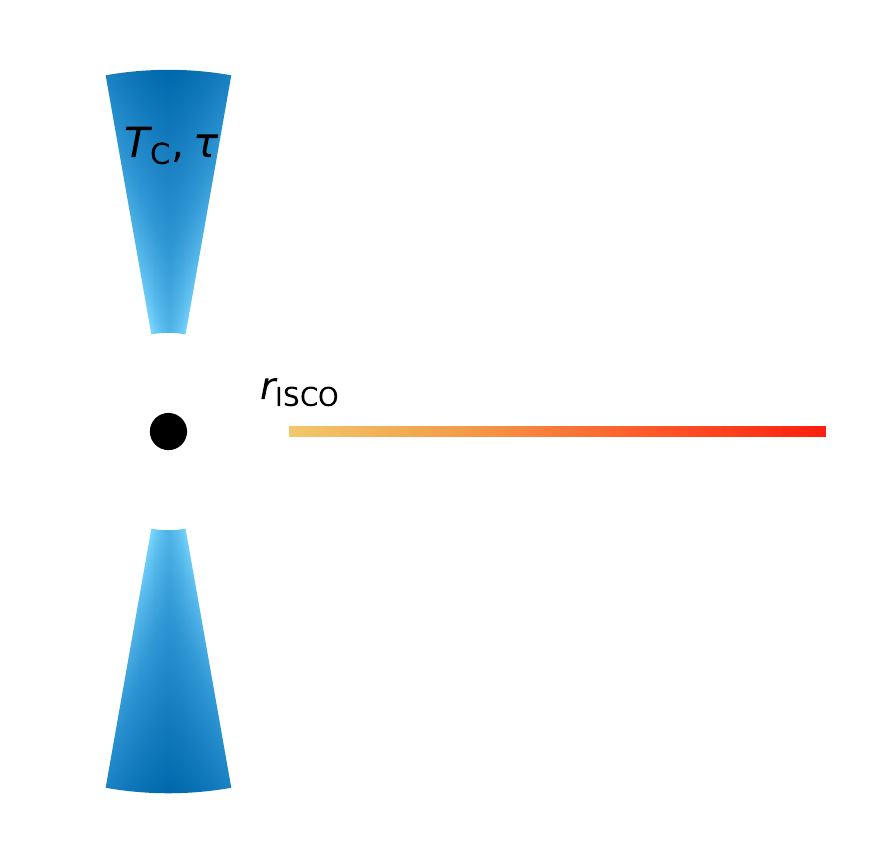}
\end{minipage}

\caption{Corona and disk geometries in \kerrc. { Left}: a wedge-shaped corona characterized by the corona opening angle $\theta_{\rm C}$, the vertical optical depth $\tauC$, and its radius $r_{\rm C}$. { Middle}: a truncated disk configuration with a wedge-shaped corona within the truncation radius $r_T$. { Right}: a lamppost-like conical corona.}
\label{fig:geometry}
\end{figure}

In \xtrack, Comptonization in the corona is modeled with a Monte Carlo scheme that uses the full relativistic Fano matrix formalism for polarized photons. In the default setup, the corona can be configured as a wedge-shaped corona sandwiching the accretion disk, a truncated disk corona, or a lamppost corona, corotating with the accretion disk at the local Keplerian velocity (see Figure \ref{fig:geometry}). The Monte Carlo framework is versatile and can accommodate arbitrary corona geometries, electron momentum distributions, and plasma bulk motions. 

For the wedge-shaped corona, we parameterize the overall scattering strength by a vertical optical depth $\tauC$, which is treated as a free parameter. During photon propagation, the differential optical depth $d\tau$ accumulated in each integration step is computed by scaling $\tauC$ by the distance traveled in that step. We emphasize that $\tauC$ parametrizes the column density of the electrons, and does not equal the actual optical depth. 

The scattering probability can be expressed as a function of the differential optical depth, such that
\begin{equation}
    P_{\rm scat}=1-e^{-d\tau}.
\end{equation}
We draw a uniform random number $p$ from $[0,1)$ after each propagation step. A scattering event is triggered if $p< P_{\rm scat}$. For each scattering event, the $k^\mu$ and $f^\mu$ vectors of the photon are first Lorentz-transformed into the local plasma comoving frame. We then sample the scattering electron's momentum direction and Lorentz factor from a Maxwell-J\"uttner distribution at the prescribed electron temperature $\TC$. The accept-reject method is used to compensate for the $(1-\beta\cos{\theta_{\rm pf}})$ factor introduced by the relative velocity effect. The $k^\mu$ and $f^\mu$ vectors are further boosted into the electron rest frame, where the polarization state is rewritten in the Stokes formalism $(I_0, Q_0, U_0)^T$. The scattering is performed in the electron rest frame by applying the relativistic Fano matrix to obtain the outgoing Stokes parameters of the scattered photon, i.e.,
\begin{equation}
\left[
\begin{array}{c}
I_1\\
Q_1 \\
U_1\\
\end{array}\right]
= \mathbf{T_{\rm F}} 
\left[
\begin{array}{c}
I_0\\
Q_0\\
U_0\\
\end{array}\right],
\end{equation}
with the Fano matrix expressed as \cite{1954PhRv...93..121F,1961RvMP...33....8M}
\begin{equation}\label{c}
\mathbf{T_{\rm F}}=\frac{1}{2}\left(\frac{x_1}{x_0}\right)^2\left[
\begin{array}{ccc}
1+\cos^2{\theta}+(x_0-x_1)(1-\cos\theta) & \sin^2\theta & 0 \\
\sin^2\theta & 1+ \cos^2\theta & 0 \\
0 & 0 & 2\cos\theta
\end{array}\right].
\end{equation}
Here 
\begin{eqnarray}
    x_1 = \frac{x_0}{1+x_0(1-\cos{\theta})}
\end{eqnarray}
is the scattered photon energy in the electron rest frame and $\theta$ is the scattering angle.

The scattered photon direction is chosen stochastically. We draw trial directions and apply another accept-reject procedure so that the accepted angles follow the required polarized Compton differential cross section distribution. Specifically, a random number $\xi$ is drawn from $[0,1)$ and the trial direction is accepted if $\xi<I_1/I_0$. 
Finally, the emergent Stokes vector is converted back into the polarization vector, and both the scattered photon momentum and polarization vectors are transformed back to the global frame for continued propagation.

It is important to note that disentangling the Comptonized component from the reflected radiation is inherently challenging in \xtrack, because the Monte Carlo approach tracks the full history of individual photons and captures different physical processes at play. For instance, a photon initially reflected by the disk may subsequently undergo Compton scattering in the corona, and conversely, a corona-scattered photon can strike the disk and be reflected back into the corona. As a result, the final emergent radiation is highly entangled, representing a mixture of reflection and Comptonization effects. Despite this entanglement, we classify the emergent photons based on whether they were reflected by the disk at any point during their propagation, regardless of their scattering histories within the corona. When generating the final results in \kerrc, these reflected photons are subsequently weighted using the \xillver reflection tables.

\begin{table*}
\centering
\caption{Simulated {\tt kerrC} black hole, accretion disk, and corona configurations. Table is adapted from Table 1 from \cite{2022ApJ...934....4K}.}
\label{table:conf}
\begin{tabularx}{\textwidth}{|p{4.2cm}|c|c|X|}
\hline
\textbf{Parameter} & \textbf{Symbol} & \textbf{Unit} & \textbf{Simulation Grid} \\
\hline
\multicolumn{4}{|c|}{\textbf{Black Hole Parameters}} \\
\hline
Black Hole Spin Parameter & $a$ & none &
$-1$, 0, 0.5, 0.75, 0.9, 0.95, 0.98, 0.99, 0.998 \\
\hline
\multicolumn{4}{|c|}{\textbf{Accretion Flow Parameters}} \\
\hline
Temperature Scaling Factor & $\sigma$ & none &
0.05, 0.1, 0.25, 0.5, 0.75, 1, 1.25, 1.5, 1.75, 2 \\
\hline
\multicolumn{4}{|c|}{\textbf{Corona Parameters}} \\
\hline
Corona Geometry & $g$ & none &
0 (wedge-shaped corona), 1 (lamppost-like conical corona), 2 (truncated disk) \\
\hline
Wedge Radius (Geom.\ \#0) & $r_C$ & $r_{\rm g}$ &
25, 50, 100 \\
\hline
Cone Start \& End (Geom.\ \#1) & $(r_1, r_2)$ & $r_{\rm g}$ &
(2.5,20), (10,50), (50,100) \\
\hline
Truncation Radius (Geom.\ \#2) & $r_{\rm T}$ & $r_{\rm g}$ &
1.237, 1.855, 2.783, 4.175, 6.264, 9.393, 14.09, 21.135, 31.703, 47.6 \\
\hline
Wedge Opening Angle (Geom.\ \#0) & $\theta_{\rm C}$ & degree &
5,\,25,\,45,\,65,\,85 \\
\hline
Cone Opening Angle (Geom.\ \#1) & $\theta_{\rm C}$ & degree &
5,\,25,\,45 \\
\hline
Corona Temperature & $T_{\rm C}$ & keV &
5, 10, 25, 100, 250, 500 \\
\hline
Corona Optical Depth & $\tau_{\rm C}$ & none &
0, 0.005, 0.014, 0.042, 0.125, 0.25, 0.5, 0.75, 1, 2 \\
\hline
\end{tabularx}
\end{table*}

\begin{figure}
    \centering
    \includegraphics[width=0.32\linewidth]{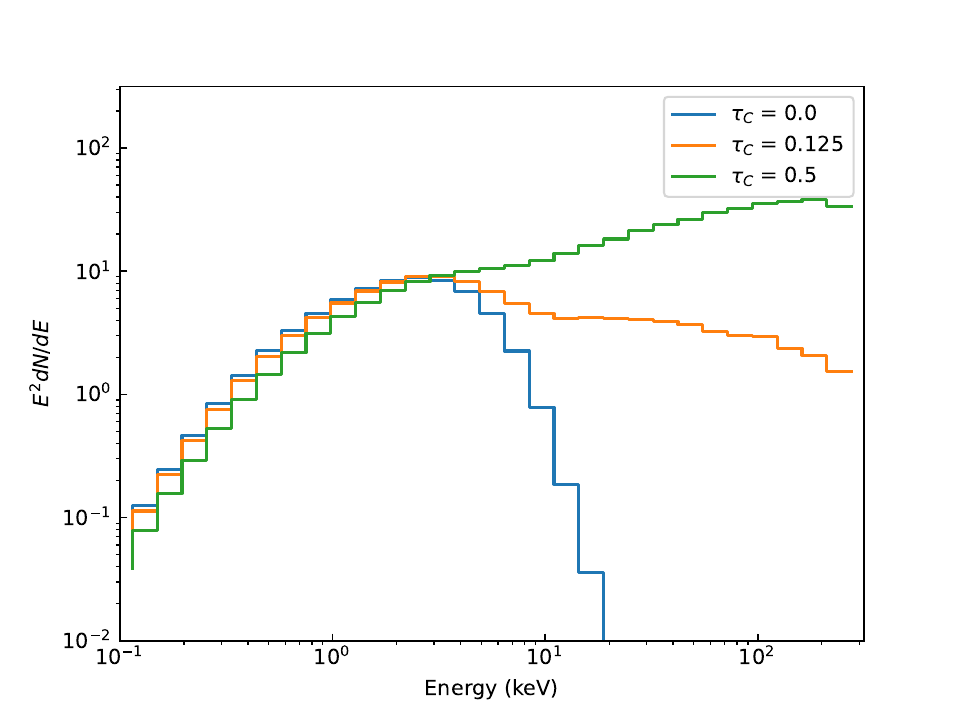}
    \includegraphics[width=0.32\linewidth]{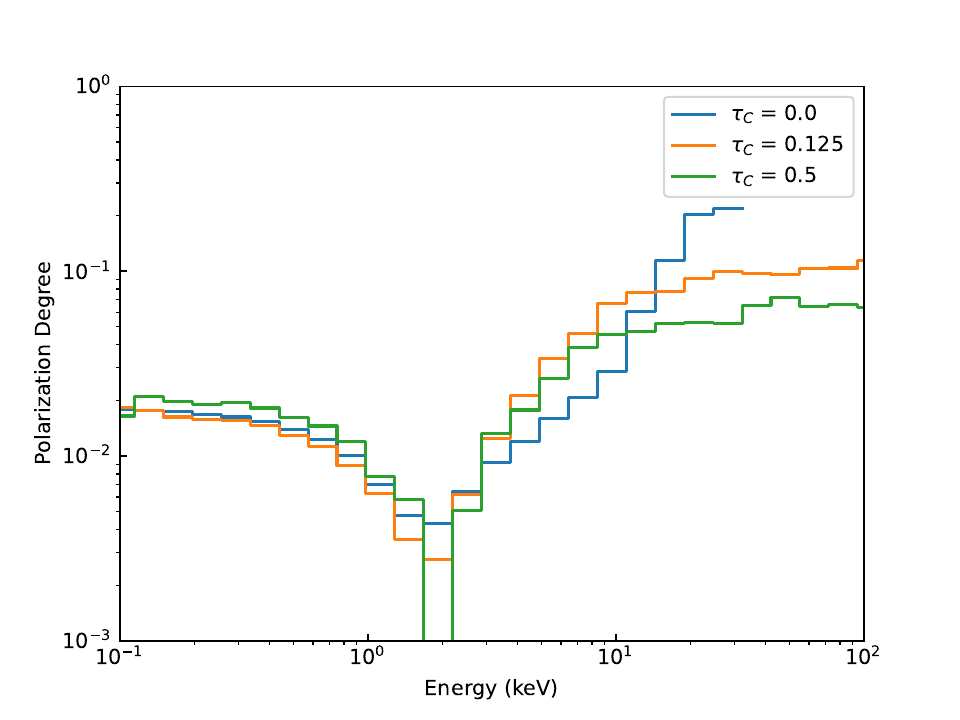}
    \includegraphics[width=0.32\linewidth]{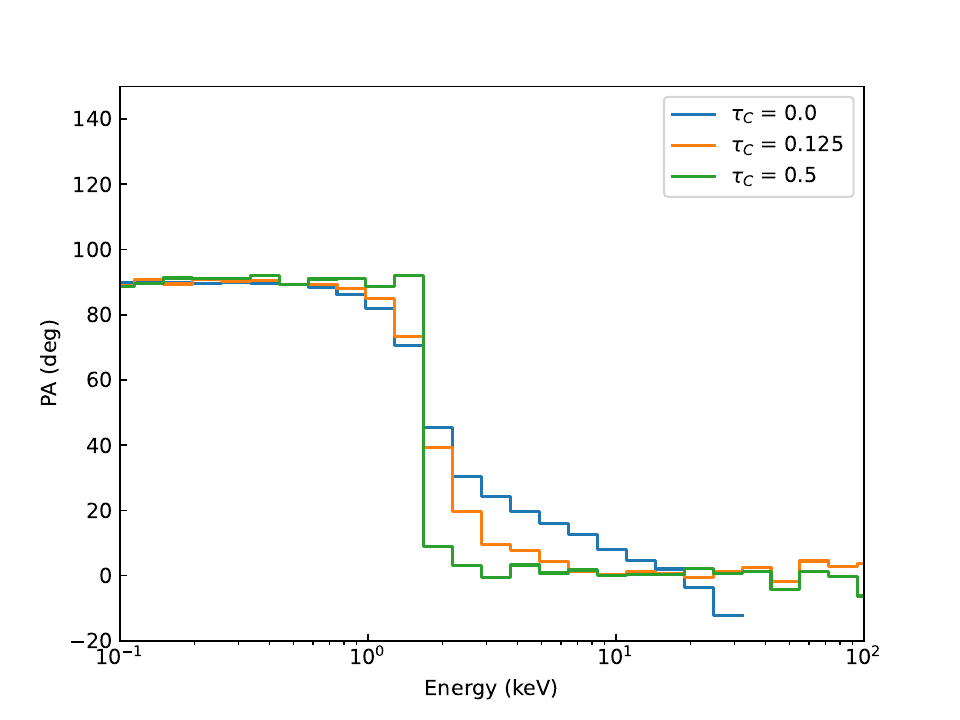}
    \caption{Simulated \xtrack spectra, polarization degrees, and polarization angles for an accreting black hole plotted as functions of energy. We assumed a black hole with a spin parameter of $a = 0.95$ and an inclination angle of 60 degrees. The temperature scaling factor $\sigma$ is fixed at 1. The color scale shows the results for different optical depths of the wedge-shaped corona with an electron temperature $kT_e = 100$ keV.}
    \label{fig:spec}
\end{figure}

The parameters of the \xtrack simulation are summarized in Table \ref{table:conf}, which includes the black hole spin $a$, the temperature scaling factor $\sigma$, and the parameters of the corona. Although the black hole mass is fixed at $10 M_{\odot}$ during the \xtrack runs, the simulated results can be scaled to arbitrary masses using the relation in Eqns.~(\ref{eqn:sigma}) and (\ref{eqn:temperature}) when fitting data with \kerrc. For the wedge-shaped corona alone, the parameter grid (7 free parameters) yields roughly $10^5$ distinct configurations. Simulations were executed for every configuration, tracking 20 million photons per run, to construct an extensive lookup table when performing spectral fitting with \kerrc.

Figure~\ref{fig:spec} illustrates the simulated spectra and polarization signals for a black hole with a wedge-shaped corona. While the $\tauC=0$ case produces a multi-temperature thermal spectrum, increasing $\tauC$ introduces a non-thermal power-law component due to the Comptonization in the hot corona. All cases exhibit a polarization angle swing near 2 keV. This feature results from the reflected emission in optically thin regimes, but is dominated by Comptonization at higher optical depths $\tauC$.

\section{KerrC Models}
\subsection{Analysis of Reflected Events}
The \kerrc fitting model post-processes the ray-traced direct disk and reflected data to generate the total spectrum. By integrating these simulated components with the \xillver tables, the model derives the energy spectrum of the partially ionized reflection. 
Specifically, the disk data are utilized to calculate the local ionization parameter 
\begin{eqnarray}
    \xi(r_i) = \frac{4\pi F(r_i)}{n_e(r_i)},
\end{eqnarray}
and the irradiation spectral index $\Gamma(r_i)$ for individual disk bins $r_i$. Here, the incident flux $F(r_i)$ is calculated using the disk data; therefore it contains the contribution from both returning disk emission and the Comptonized corona emission. These localized parameters are then used to extract the correct reflection flux 
\begin{eqnarray}
    F_X = F_X(\Gamma,A_{Fe},\log_{10}\xi,kT_{\rm e},\log_{10}n_{\rm e},\theta;E)
\end{eqnarray} 
from the XILLVER atomic table {\tt xillverCp\_v3.6.fits}. Here $A_{Fe}$ is the relative iron abundance, $kT_e$ is the electron temperature, $n_e$ is the electron density, $\theta$ is the inclination angle of the reflected radiation in the plasma frame and $E$ is the photon energy. 
A weight ratio
\begin{equation}
    r\,=\,\frac
    {F_X(\Gamma_i,A_{Fe},\xi(r_i),kT_{\rm e},
    n_{\rm e}(r_i),\theta;E)}
    {F_X(\Gamma_i,A_{Fe},\xi_{\rm max},kT_{\rm e},
    n_{\rm e}(r_i),\theta;E)}
    \label{eqn:weight_ratio}
\end{equation}
is subsequently applied to the reflected emission data to construct the final, physically rigorous spectrum. The denominator of Eqn.~(\ref{eqn:weight_ratio}) gives the energy flux with the highest ionization parameter ($\log_{10}\xi_{\rm max}$=4.7) in the \xillver table, which agrees with the Chandrasekhar's model for a fully ionized electron atmosphere. 
Note that the factor $r$ modifies the intensity of the reflected emission relative to the result obtained with Chandrasekhar's formalism, and it will be applied to each photon event in the reflected emission data to obtain the physical reflected flux. The polarization of the reflected emission is assumed to be given by Chandrasekhar's results as well.

\subsection{KerrC and KerrClight Fitting Models}

The \kerrc model is a spectral-fitting tool that leverages a vast library of HDF5 emission datasets generated by the \xtrack ray-tracing code. It is designed for seamless integration and can be easily loaded as a user-defined model with the {\tt Sherpa} package. As outlined in Table \ref{table:par}, {\tt kerrC} provides a wide range of physical parameters. These include the black hole properties (such as the mass, spin, distance, inclination, and accretion rate); the corona parameters (including corona size, temperature and optical depth); and the reflection parameters (such as metallicity, iron abundance, electron temperature, and electron density). The model also allows users to choose between distinct spatial corona geometries: a standard wedge corona, a lamppost corona, or a wedge corona paired with a truncated accretion disk.

\begin{table}[t!]
\centering
\caption{{\tt kerrc} model parameters. Adapted from Table 2 of \cite{2022ApJ...934....4K}.}
\label{table:par}
\begin{tabularx}{\linewidth}{|c|>{\raggedright\arraybackslash}X|c|c|c|>{\raggedright\arraybackslash}X|}
\hline
\textbf{Number} & \textbf{Parameter} & \textbf{Symbol} & \textbf{\kerrc name} & \textbf{Unit} & \textbf{Allowed Values} \\ \hline
\multicolumn{6}{|c|}{\textbf{Black Hole Parameters}}\\ \hline
1 & Black Hole Mass & $M$ & {\tt M} & $M_{\odot}$ & any \\
2 & Black Hole Spin Parameter & $a$ & {\tt a} & none & -1 ... 0.998 \\
3 & Black Hole Inclination & $i$ & {\tt incl} & degree & 5 ... 85 \\
4 & Distance & $d$ & {\tt dist} & kpc & 0 ...  \\ \hline
\multicolumn{6}{|c|}{\textbf{Accretion Flow Parameters}}\\ \hline
5 & Mass Accretion Rate & $\dot{M}$ & {\tt mDot} & $10^{18} \,{\rm g\, s}^{-1}$ & depends on $M$ \\ \hline
\multicolumn{6}{|c|}{\textbf{Corona Parameters}}\\ \hline
6 & Corona Geometry & $g$ & {\tt geom} & none & 0 (wedge), 1 (lamppost), 2 (truncated disk) \\
7 & Corona Edge & $r_{\rm C}$ & {\tt rC} & $r_{\rm g}$ & 2.5 ... 100 \\
8 & Corona Opening Angle & $\theta_{\rm C}$ & {\tt thetaC}& degree & 5 ... 85 \\
9 & Corona Temperature & $kT_{\rm C}$ & {\tt tempC} & keV & 5 ... 500 \\
10& Corona Optical Depth & $\tau_{\rm C}$ & {\tt tauC} & none & 0 ... 2 \\ \hline
\multicolumn{6}{|c|}{\textbf{Reflection Parameters}}\\ \hline
11& Reflection Amplitude & - & {\tt refl} & none & 0 ... 10 \\ 
12& Metallicity & $A_{Fe}$ &  {\tt AFe} &  solar & 0.5 ... 10\\
13& Electron Temperature & $kT_{\rm e}$ & {\tt kTe}&keV & 1 ... 400 \\
14& Electron Density at $r_{\rm ISCO}$& log$_{10}\,n_{\rm e,0}$ & {\tt logDens}  & cgs & 15 ... 22\\ 
15& Radial Density Power-Law Index & $\alpha$ & {\tt xiIndex} & none & 0 ...  \\ 
16& $\Gamma$-fit Range & $\kappa$ & {\tt kappa} & none & 0.1 ... 10 \\ \hline 
\multicolumn{6}{|c|}{\textbf{Polarization Parameters}}\\ \hline
17& Angle of BH Spin Axis from Celestial North Pole, Counter-Clockwise & $\chi$ & {\tt chi} & degree & 0 ... 180 \\ \hline
\multicolumn{6}{|c|}{\textbf{Additional Model Modifiers}}\\ \hline
18& Inner Cutoff  & $l_1$ & {\tt l1} & $r_{\rm g}$ & 0, $r_{\rm ISCO}$ ... 100 \\
19& Outer Cutoff  & $l_2$ & {\tt l2} & $r_{\rm g}$ & 0, $r_{\rm ISCO}$ ... 100 \\ \hline
\end{tabularx}

\vspace{2ex}
\raggedright
\end{table}

When the \kerrc model is invoked in \sherpa with a specific set of parameters, it calculates the Stokes $I$, $Q$, and $U$ spectra using simplex interpolation. Because the ray-tracing model operates with 6 free parameters, namely, black hole spin, temperature scaling factor, corona temperature, corona optical depth, corona opening angle, and corona size, the simplex algorithm efficiently interpolates across the high-dimensional parameter space using the 7 nearest simulation runs from the database. The rest of the parameters in Table \ref{table:par} are not part of the ray-tracing grid, and they are adjusted via rescaling during the fitting process. The resulting $I$, $Q$, and $U$ spectra can then be folded through instrumental responses and combined with absorption models or other model components to fit observational X-ray data. 

In addition to the standard \kerrc model, we also provide {\tt kerrClight}, a simplified and optimized version of \kerrc. 
Rather than utilizing the reflection table for the partially ionized disk, {\tt kerrClight} approximates the disk reflection using Chandrasekhar's fully ionized reflection description \cite{2024ApJ...977L..10K}.
Because this approach bypasses the computationally heavy post-processing of the disk reflected data, spectral fitting with {\tt kerrClight} is significantly faster, making it a perfect tool for rapid parameter space exploration and preliminary fitting.


\section{Application of KerrC Modeling}
\subsection{Understanding the Coronal Geometry and Spin of Stellar Mass Black Holes}
\label{}
The \kerrc and {\tt kerrClight} are highly advanced simulation models designed to extract the geometry and spin of black hole X-ray binaries from combined spectral and polarimetric observations. Their primary strength lies in their ability to treat the direct disk emission, the disk reflection, and the GR ray-tracing self-consistently. The simulation captures the full suite of GR effects, including light bending, GR redshift and parallel transport of the polarization 4-vectors. Therefore, it allows an accurate map of the physical environments near the event horizon to the spectral and polarization signal in the X-ray band.

\begin{figure}
    \centering
\resizebox{1.15\textwidth}{!}{%
    \includegraphics[height=11cm]{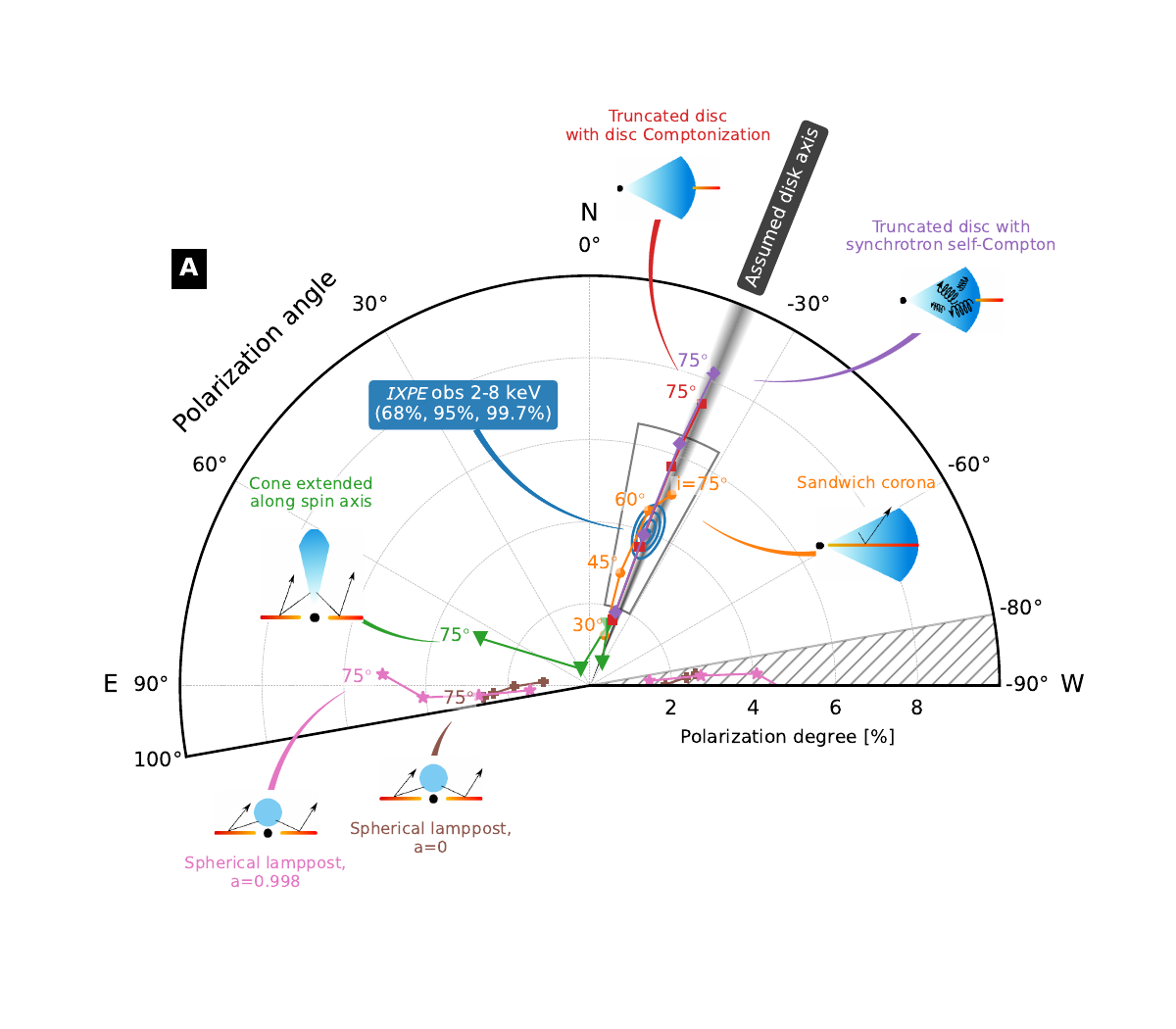}%
    \hspace*{-1cm}%
    \includegraphics[trim={5.8cm 0 0 0},clip,height=11cm]{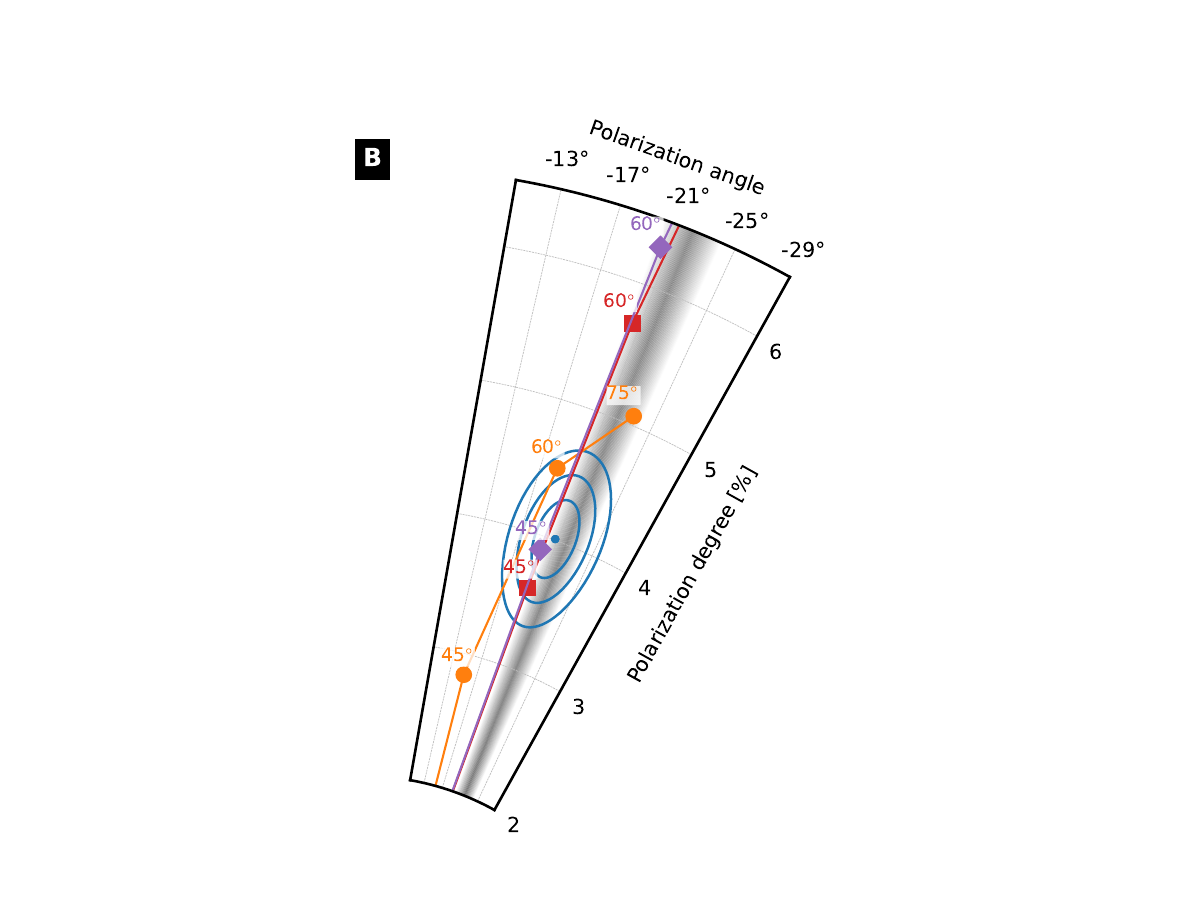}%
}
    \caption{Comparison of the Cyg X-1 hard state \emph{IXPE} polarization measurements in the 2--8 keV band with model predictions reprinted from \cite{2022Sci...378..650K} with permission. The blue dots in the plot display the observed polarization degree and polarization angle, and the blue ellipses indicate the $1\sigma$, $2\sigma$, and $3\sigma$ confidence levels. The gray band displays the orientation of the radio jet. The connected data points with different colors represent the polarization predictions from different corona models with different inclination angles. Models with a corona extending parallel to the accretion disk match the observed polarization. However, models with a corona extending perpendicular to the disk plane predict polarization direction parallel to the disk plane are therefore disfavored by the \emph{IXPE} results. Panel B provides a close-up view of the region corresponding to the \emph{IXPE} measurements.}
    \label{fig:cyg_x1_hard_state}
\end{figure}

For black hole X-ray binaries in the low hard state, \kerrc has proven essential for helping to resolve long-standing debates about the shape and location of the hot corona. For example, Krawczynski et al. (2022) utilized the {\tt kerrC} model to analyze the polarization data from \emph{IXPE} alongside multiband spectral data for the benchmark black hole binary Cygnus X-1 \cite{2022Sci...378..650K}. The observation revealed a significant polarization degree of around 4\% in the 2--8 keV band, with the polarization direction closely aligned with the system's radio jet. The {\tt kerrC} model indicates that the observed polarization signal can be achieved in a wedge- or sandwich-shaped corona. Since scatterings preserve polarization perpendicular to the scattering plane, multiple scatterings in a horizontal extended corona in the plane of the accretion disk would build up polarization perpendicular to the accretion disk (see Figure \ref{fig:cyg_x1_hard_state}). In contrast,  a ``lamppost" or vertically extended corona along the jet axis generates polarization parallel to the disk plane. This horizontally extended corona geometry is further supported by the \emph{XL-Calibur} hard X-ray polarization observations, which measured a polarization fraction of approximately 5\% in Cyg X-1, with the polarization angle aligned with the jet direction \cite{2025ApJ...994...37A}.

\begin{figure}
    \centering
    \includegraphics[width=0.45\linewidth]{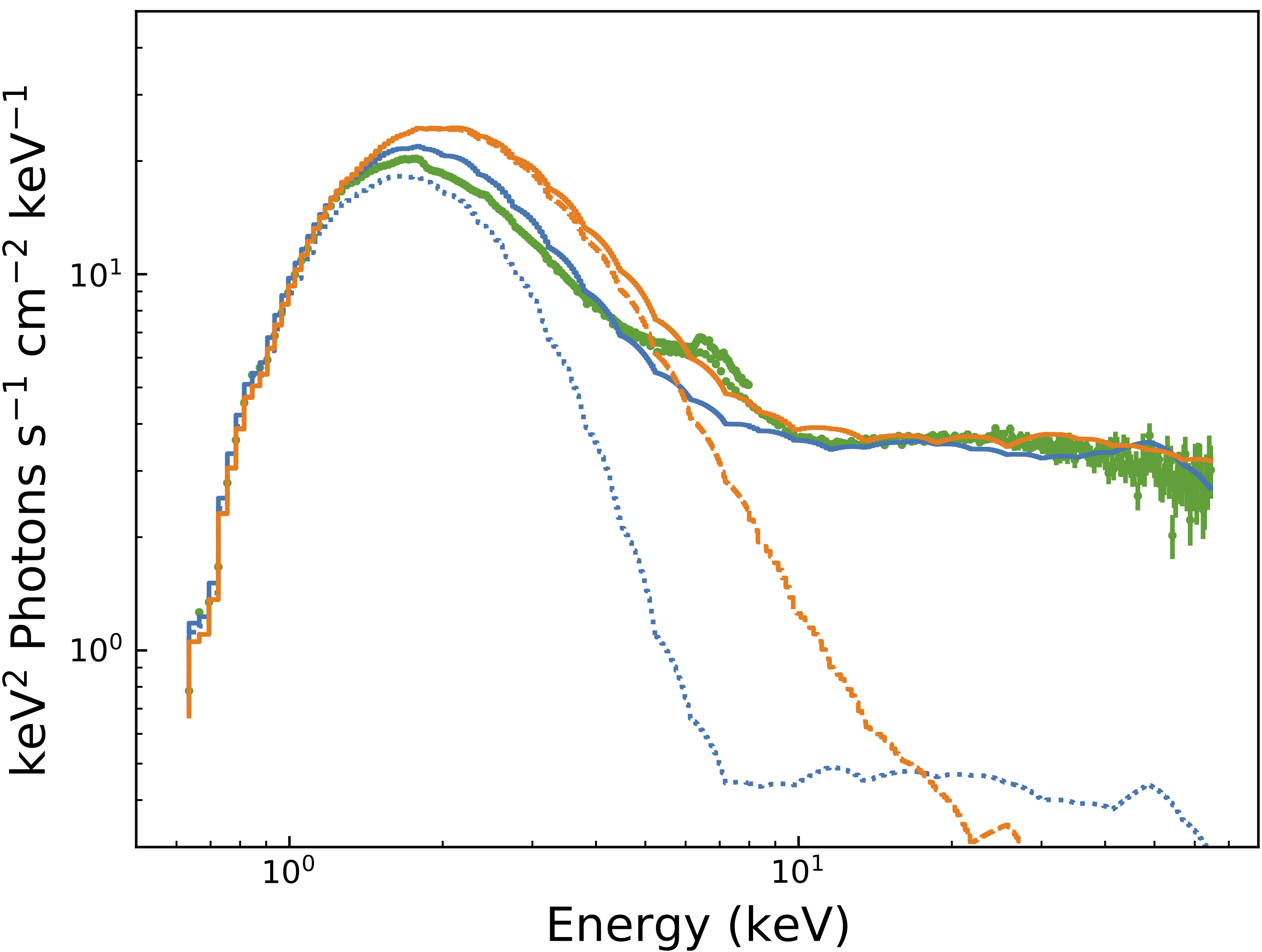}
     \includegraphics[width=0.45\linewidth]{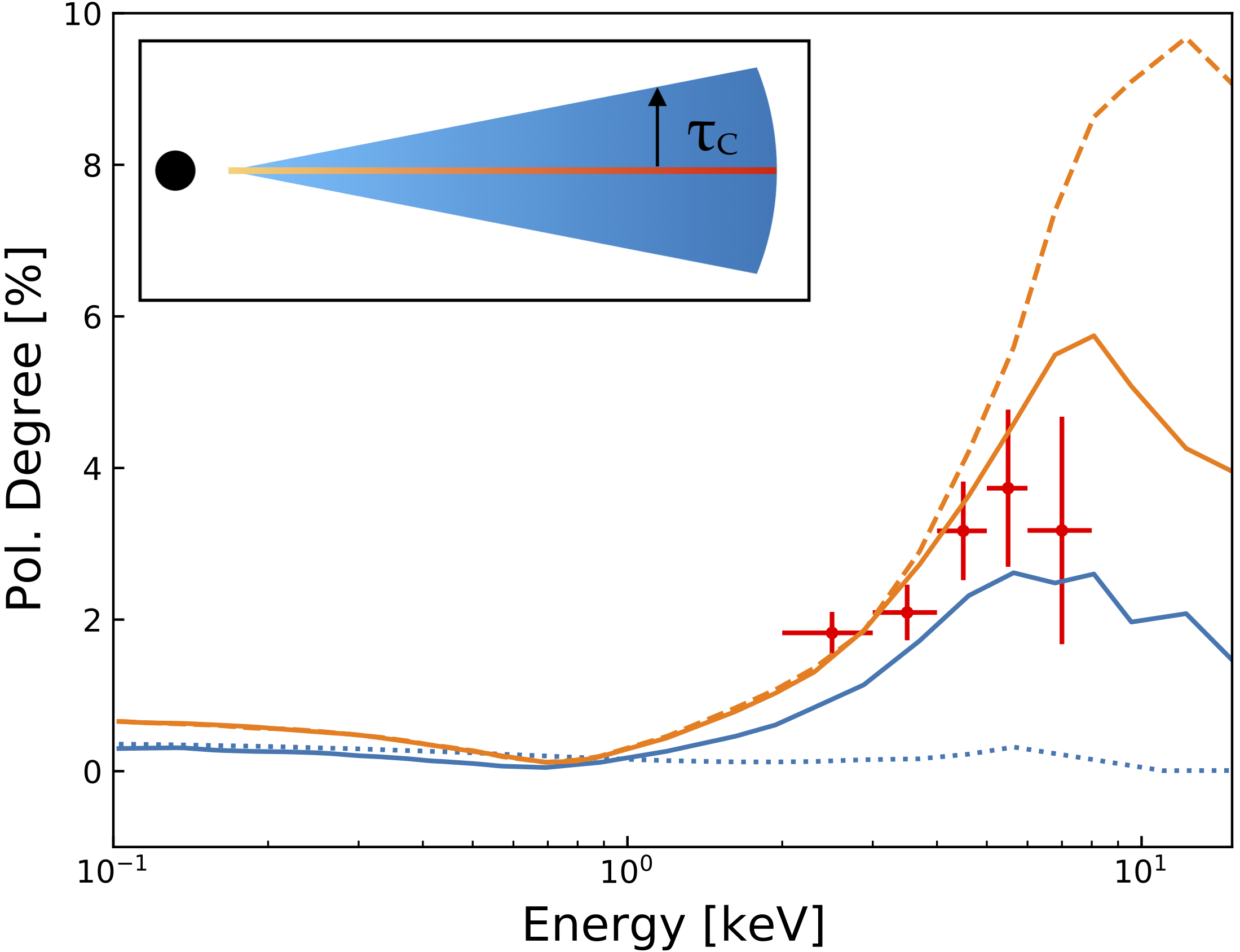}
    \caption{Simulated spectrum and polarization degree from the \kerrc model superimposed with the observational data for Cyg X-1 \cite{2024ApJ...969L..30S}. The green data points in the left panel are from \emph{NICER} and \emph{NuSTAR}, and the red data points display the simultaneous \emph{IXPE} polarization measurements. In both panels, the solid orange curves represent a model with an inclination angle of $40^\circ$ and an optical depth of $\tauC=0.007$, while the dashed orange curves show the corresponding results for $\tauC=0$. The blue curves represent models with an inclination angle of $27^\circ$; the solid blue curves correspond to $\tauC=0.007$, and the dotted blue curves show the results for direct emission only.}
    \label{fig:cyg_x1_soft_state}
\end{figure}

Beyond the hard state, \kerrc models have been used to decipher the soft state of X-ray binaries, where the thermal disk emission dominates the X-ray spectrum. An example is the \emph{IXPE} campaign on the soft state of Cygnus X-1 \cite{2024ApJ...969L..30S}. In the soft state, \emph{IXPE} detected a roughly 2\% polarization degree, which remained aligned with the radio jet. This is in contrast to the classical picture, where the thermal emission from the disk is polarized parallel to the disk. The {\tt kerrC} models revealed that this polarization signature could be explained by the returning radiation, which is originally emitted by the inner disk, but bent backward by gravity, and reflected off the disk surface before escaping to the observer (see Figure \ref{fig:cyg_x1_soft_state}). The polarization direction is also distorted by the strong gravity into a direction almost perpendicular to the disk. In order to achieve this strong gravitational lensing effect, the black hole's spin parameter has to be near maximal $(a>0.96)$.

\subsection{Future Developments}

While the \kerrc model provides an essential physical tool to study the emission mechanism and geometries of black hole accretion systems, future improvements can be made for both the underlying physical modeling and the computational algorithms used in data fitting. On the physical front, future upgrades will include transitioning the corona configuration to a dynamical, outflowing model. This advancement will allow \kerrc to explore the polarization enhancement driven by coronal outflow, offering a natural explanation for the observed large polarization degrees without requiring large viewing inclinations or maximal black hole spin \cite{2001MNRAS.326..417M,2023ApJ...949L..10P,2025ApJ...993...54K}. Additionally, we will update the ray-tracing simulations to incorporate a self-consistent reflection feedback mechanism \cite{1991ApJ...380L..51H}. By accounting for the heating caused by the emission coming back to the disk, this update will allow us to calculate a more physically motivated temperature profile for the accretion disk. 

In addition to these physical improvements, the standard method of interpolating across discrete ray-tracing data grids will be upgraded using machine learning techniques. By training neural networks to learn the non-linear, continuous mappings between physical parameter inputs and the output Stokes spectra, we can bypass the inaccuracies of the edge-boundary grid interpolation, accelerating and refining the parameter estimation process.



\begin{acknowledgement}
KH and HK thank NASA for support
under the grants 
80NSSC24K0205, 80NSSC24K1178, 80NSSC24K1819, and 80NSSC24K1749, as well as from the McDonnell Center for the Space Sciences.
\end{acknowledgement}


\begin{thebibliography}{99}































































\bibitem{2021APh...12602529A}
Q. Abarr, H. Awaki, M.~G. Baring, R. Bose, G. De Geronimo, P. Dowkontt, M. Errando, V. Guarino, K. Hattori, K. Hayashida, F. Imazato, M. Ishida, N.~K. Iyer, F. Kislat, M. Kiss, T. Kitaguchi, H. Krawczynski, L. Lisalda, H. Matake, Y. Maeda, H. Matsumoto, T. Mineta, T. Miyazawa, T. Mizuno, T. Okajima, M. Pearce, B.~F. Rauch, F. Ryde, C. Shreves, S. Spooner, T.-A. Stana, H. Takahashi, M. Takeo, T. Tamagawa, K. Tamura, H. Tsunemi, N. Uchida, Y. Uchida, A.~T. West, E.~A. Wulf, \& R. Yamamoto,
Astroparticle Physics \textbf{126}, 102529 (2021)\ doi:10.1016/j.astropartphys.2020.102529

\bibitem{2025ApJ...994...37A}
H. Awaki, M.~G. Baring, R. Bose, J. Casey, S. Chun, A. Dasgupta, P. Galchenko, E. Gau, K. Goya, T. Hakamata, T. Hayashi, S. Heatwole, K. Hu, D. Ishi, M. Ishida, F. Kislat, M. Kiss, K. Klepper, H. Krawczynski, H. Kuramoto, L. Lisalda, Y. Maeda, H. Matsumoto, S.~V. Menon, A. Miyamoto, A. Miyamoto, K. Murakami, T. Okajima, M. Pearce, B. Rauch, N. Rodriguez Cavero, K. Shirahama, S. Spooner, H. Takahashi, K. Tamura, Y. Uchida, K. Wimalasena, M. Yokota, \& M. Yoshimoto,
The Astrophysical Journal \textbf{994}, 37 (2025)\ doi:10.3847/1538-4357/ae0f1d

\bibitem{Cash:1990jdx}
J.~R.~Cash and A.~H.~Karp,
ACM Trans. Math. Software \textbf{16}, no.3, 201-222 (1990)
doi:10.1145/79505.79507

\bibitem{1960ratr.book.....C}
S. Chandrasekhar,
New York: Dover (1960)

\bibitem{1977Natur.269..128C}
P.~A. Connors and R.~F. Stark,
Nature \textbf{269}, 128-129 (1977)
doi:10.1038/269128a0

\bibitem{2025ApJ...995L..12D}
P.~A. Draghis, J.~M. Miller, E. Kara, E. Costantini, O. Adegoke, \& J.~A. Garc{\'\i}a,
The Astrophysical Journal \textbf{995}, L12 (2025)\ doi:10.3847/2041-8213/ae2276

\bibitem{1954PhRv...93..121F}
U. Fano,
Physical Review \textbf{93}, 121-123 (1954)
doi:10.1103/PhysRev.93.121

\bibitem{2004MNRAS.355.1105F}
R.~P. Fender, T.~M. Belloni, \& E. Gallo,
Monthly Notices of the Royal Astronomical Society \textbf{355}, 1105 (2004)\ doi:10.1111/j.1365-2966.2004.08384.x

\bibitem{2010ApJ...718..695G}
J. Garc{\'\i}a \& T.~R. Kallman,
The Astrophysical Journal \textbf{718}, 695 (2010)\ doi:10.1088/0004-637X/718/2/695

\bibitem{2013ApJ...768..146G}
J. Garc{\'\i}a, T. Dauser, C.~S. Reynolds, T.~R. Kallman, J.~E. McClintock, J. Wilms, \& W. Eikmann,
The Astrophysical Journal \textbf{768}, 146 (2013)\ doi:10.1088/0004-637X/768/2/146

\bibitem{2014ApJ...782...76G}
J. Garc{\'\i}a, T. Dauser, A. Lohfink, T.~R. Kallman, J.~F. Steiner, J.~E. McClintock, L. Brenneman, J. Wilms, W. Eikmann, C.~S. Reynolds, \& F. Tombesi,
The Astrophysical Journal \textbf{782}, 76 (2014)\ doi:10.1088/0004-637X/782/2/76

\bibitem{2016SPIE.9905E..1HG}
K.~C. Gendreau, Z. Arzoumanian, et al.,
Space Telescopes and Instrumentation 2016: Ultraviolet to Gamma Ray \textbf{9905}, 99051H (2016)\ doi:10.1117/12.2231304

\bibitem{1991ApJ...380L..51H}
F. Haardt \& L. Maraschi,
The Astrophysical Journal \textbf{380}, L51 (1991)\ doi:10.1086/186171

\bibitem{2013ApJ...770..103H}
F.~A. Harrison, W.~W. Craig, et al.,
The Astrophysical Journal \textbf{770}, 103 (2013)\ doi:10.1088/0004-637X/770/2/103

\bibitem{2001A&A...365L...1J}
F. Jansen, D. Lumb, et al.,
Astronomy and Astrophysics \textbf{365}, L1 (2001)\ doi:10.1051/0004-6361:20000036

\bibitem{2019ApJ...873...71K}
B.~E. Kinch, J.~D. Schnittman, T.~R. Kallman, \& J.~H. Krolik,
The Astrophysical Journal \textbf{873}, 71 (2019)\ doi:10.3847/1538-4357/ab05d5

\bibitem{2012ApJ...754..133K}
H. Krawczynski,
The Astrophysical Journal \textbf{754}, 133 (2012)\ doi:10.1088/0004-637X/754/2/133

\bibitem{2022ApJ...934....4K}
H. Krawczynski \& B. Beheshtipour,
The Astrophysical Journal \textbf{934}, 4 (2022)\ doi:10.3847/1538-4357/ac7725

\bibitem{2025ApJ...993...54K}
H. Krawczynski \& K. Hu,
The Astrophysical Journal \textbf{993}, 54 (2025)\ doi:10.3847/1538-4357/adfddf

\bibitem{2022Sci...378..650K}
H. Krawczynski, F. Muleri, et al.,
Science \textbf{378}, 650 (2022)\ doi:10.1126/science.add5399

\bibitem{2024ApJ...977L..10K}
H. Krawczynski, Y. Yuan, et al.,
The Astrophysical Journal \textbf{977}, L10 (2024)\ doi:10.3847/2041-8213/ad855c

\bibitem{2005ApJS..157..335L}
L.-X. Li, E.~R. Zimmerman, R. Narayan, \& J.~E. McClintock,
The Astrophysical Journal Supplement Series \textbf{157}, 335 (2005)\ doi:10.1086/428089

\bibitem{2001MNRAS.326..417M}
J. Malzac, A.~M. Beloborodov, \& J. Poutanen,
Monthly Notices of the Royal Astronomical Society \textbf{326}, 417 (2001)\ doi:10.1046/j.1365-8711.2001.04450.x

\bibitem{1961RvMP...33....8M}
W.~H. McMaster,
Reviews of Modern Physics \textbf{33}, 8-27 (1961)
doi:10.1103/RevModPhys.33.8

\bibitem{2026arXiv260103349N}
C. Nagele, J.~H. Krolik, R. Liu, B.~E. Kinch, \& J.~D. Schnittman,
arXiv e-prints arXiv:2601.03349 (2026)\ doi:10.48550/arXiv.2601.03349

\bibitem{Novikov:1973kta}
I.~D. Novikov and K.~S. Thorne,
Black Holes (Les Astres Occlus), 343–450 (1973)

\bibitem{Page:1974he}
D.~N. Page and K.~S. Thorne,
Astrophys. J. \textbf{191}, 499-506 (1974)
doi:10.1086/152990

\bibitem{1996ApJ...470..249P}
J. Poutanen and R. Svensson,
The Astrophysical Journal \textbf{470}, 249 (1996)
doi:10.1086/177865

\bibitem{2023ApJ...949L..10P}
J. Poutanen, A. Veledina, \& A.~M. Beloborodov,
The Astrophysical Journal \textbf{949}, L10 (2023)\ doi:10.3847/2041-8213/acd33e

\bibitem{1983ASPRv...2..189P}
L.~A. Pozdnyakov, I.~M. Sobol, and R.~A. Syunyaev,
Astrophysics and Space Physics Reviews \textbf{2}, 189-331 (1983)

\bibitem{2012ApJ...745....1P}
D.\ Psaltis, and T.\ Johannsen,
The Astrophysical Journal \textbf{745}, 1 (2012)\ doi:10.1088/0004-637X/745/1/1

\bibitem{2006ARA&A..44...49R}
R.~A. Remillard \& J.~E. McClintock,
Annual Review of Astronomy and Astrophysics \textbf{44}, 49 (2006)\ doi:10.1146/annurev.astro.44.051905.092532

\bibitem{2009ApJ...701.1175S}
J.~D. Schnittman \& J.~H. Krolik,
The Astrophysical Journal \textbf{701}, 1175 (2009)\ doi:10.1088/0004-637X/701/2/1175

\bibitem{2013ApJ...777...11S}
J.~D. Schnittman \& J.~H. Krolik,
The Astrophysical Journal \textbf{777}, 11 (2013)\ doi:10.1088/0004-637X/777/1/11

\bibitem{Shakura:1972te}
N.~I. Shakura and R.~A. Sunyaev,
Astron. Astrophys. \textbf{24}, 337-355 (1973)

\bibitem{1995ApJ...445..780S}
T. Shimura and F. Takahara,
The Astrophysical Journal \textbf{445}, 780 (1995)
doi:10.1086/175740

\bibitem{2024ApJ...969L..30S}
J.~F. Steiner, E. Nathan, et al.,
The Astrophysical Journal \textbf{969}, L30 (2024)\ doi:10.3847/2041-8213/ad58e4

\bibitem{1980A&A....86..121S}
R.~A. Sunyaev \& L.~G. Titarchuk,
Astronomy and Astrophysics \textbf{86}, 121 (1980)

\bibitem{2021MNRAS.501.3393T}
R. Taverna, L. Marra, et al.,
Monthly Notices of the Royal Astronomical Society \textbf{501}, 3393 (2021)\ doi:10.1093/mnras/staa3859

\bibitem{2024icrc.confE.745T}
J. Tomsick, S. Boggs, et al.,
38th International Cosmic Ray Conference \textbf{745} (2024)\ doi:10.22323/1.444.0745

\bibitem{Walker:1970un}
M.~Walker and R.~Penrose,
Commun. Math. Phys. \textbf{18}, 265-274 (1970)
doi:10.1007/BF01649445

\bibitem{2022JATIS...8b6002W}
M.~C. Weisskopf, P. Soffitta, et al.,
Journal of Astronomical Telescopes, Instruments, and Systems \textbf{8}, 026002 (2022)\ doi:10.1117/1.JATIS.8.2.026002

\bibitem{2020arXiv200304962X}
XRISM Science Team,
arXiv e-prints arXiv:2003.04962 (2020)

\bibitem{2025PASJ...77.1210Y}
S. Yamada, N. Hell, et al.,
Publications of the Astronomical Society of Japan \textbf{77}, 1210 (2025)\ doi:10.1093/pasj/psaf104

\bibitem{1996MNRAS.283..193Z}
A.~A. Zdziarski, W.~N. Johnson, \& P. Magdziarz,
Monthly Notices of the Royal Astronomical Society \textbf{283}, 193 (1996)\ doi:10.1093/mnras/283.1.193

\bibitem{2019ApJ...875..148Z}
W. Zhang, M. Dov\v{c}iak, \& M. Bursa,
The Astrophysical Journal \textbf{875}, 148 (2019)\ doi:10.3847/1538-4357/ab1261


\end{thebibliography}
\end{document}